\documentclass[preprint]{revtex4-1}
\usepackage{graphicx}
\usepackage{dcolumn}
\usepackage{amsthm}
\usepackage{amssymb}
\newtheorem{mydef}{Definition}

\begin{document}
\title{On the Physical Cluster of Nucleation Theory: Generalized Hill Cluster}
\author{Rasmus A. X. Persson}
\email{rasmus.persson@chem.gu.se}
\affiliation{Department of Chemistry, University of Gothenburg, Sweden}

\begin{abstract}
The physical cluster originally conceived by Hill (J. Chem. Phys, 23, 617) is
generalized for the case $N > 2$ in a novel way. Contrary to Hill's pairwise
generalization, this definition assures that all constituent molecules of the
cluster have insufficient kinetic energy to escape, and avoids the spurious
implication of Hill's generalization that the molecular velocities are to be
increasingly equal.
\end{abstract}

\maketitle

\section{Introduction}
The rate of a phase transition obviously depends to a large extent on the rate
at which the molecules can rearrange themselves, and in the case of
condensation especially, this rate is much smaller than the typical
rate of molecular translation. A condensation process necessitates
diffusion along the concentration gradient and would be thermodynamically
forbidden were it not for the decrease of chemical potential associated with
the transformation, due to the attractive interactions dominating in the
condensed phase. However, in the very early stage of the phase transformation,
insufficient condensed phase is present to afford this stabilization. In these
cases, the transport of matter must occur along the chemical potential
gradient. The great reluctancy by which this fluctuation takes place is one
bottleneck in the condensation process and was first identified by
Gibbs, who expressed the free energy of a small cluster as the sum of a surface
and a bulk term, taking the surface term to be proportional to the specific
surface free energy of the bulk liquid \cite{gibbs28}. This approach was
further extended into a theory for the kinetics of the process by a number of
authors \cite{becker35,frenkel39,zeldovich43}. This classical nucleation
theory, as it has since become known, predicts a nucleation rate in fortuitous
accord with experiment for water condensation in a narrow temperature interval.

In efforts to side track the assumptions of bulk properties for small clusters
on the order of 50-100 monomers, much computer simulation, following the
pioneering work of Lee {\em et al.} \cite{lee73}, has been carried out over the
years, and new theories developed. However, the concept of a cluster has in
these simulations often been quite arbitrary. Most often based on a simple
distance criterion, as in {\em e. g.} the Stillinger \cite{stillinger63} or LBA
\cite{lee73} cluster, there is no compelling argument {\em a priori} for the
particular dimension of the cluster. The same problem of arbitrariness is faced
when exchanging the distance criterion for one of potential energy
\cite{chen03}.

Of particular elegance is then the definition due to Hill \cite{hill55}, who
regarded molecules in regions of phase space of negative total energy to form a
cluster, in analogy with how ``bound states'' are usually defined in the rest
of physics, for instance between atoms in molecules, or between celestial
bodies and satellites in astronomy. Moreover, the concept of the critical
temperature follows directly from this notion, as the effective temperature
where the kinetic energy is so great that no bound states can be formed, even
for the infinite cluster, {\em i. e.} bulk.  Following Hill, we denote such a
cluster, a ``physical cluster.'' However, Hill's generalization of his cluster
definition past the dimer is based on a pairwise expansion in which monomers
are not bound to the cluster as a whole, but to each other, and while rigorous,
is not completely intuitive. In fact, for large clusters, his definition
entails that the velocities of the monomers become increasingly
harmonized \footnote{This follows because there can be no collective binding in
the Hill cluster, therefore the potential well in which the molecule is
considered trapped does not increase with the size of the cluster, and
consequently, the relative velocities must remain low.}. In the
author's view, nevertheless, one of the most important concepts of the Hill
cluster is its inherent non-locality. Such non-locality has been exploited in
other theories of nucleation, the most prominent being the density-functional
approach of Oxtoby and Evans \cite{oxtoby88}.

Consider the set $\mathbb D_N$ of $N$ particles ordered from $1$ to $N$. The
particles have positions $\{\vec r_i\}_1^N$, velocities $\{\vec v_i\}_1^N$ and
masses $\{m_i\}_1^N$. Let $U(\{\vec r_i\})$ denote the potential energy of the
system and let $U(\{\vec r_i\})$ have the property that $U(\{\vec r_i\}) \to
0$, when the smallest mutual distance in $\{\vec r_i\}$ approaches positive
infinity. Let us first recall Hill's definition of the physical dimer.
\begin{mydef}
\label{def:dimer}
The two particles $i$ and $j$ constitute a physical dimer if $\frac 1 2 \mu
(\vec v_i - \vec v_j)^2 \leq -U(\vec r_i, \vec r_j)$ where $\mu = \frac {m_i
m_j} {m_i + m_j}$.
\end{mydef}

Let us now define two ordered subsets of $\mathbb D_N$ denoted by $\mathbb A$
and $\mathbb B$ such that $\mathbb A \cup \mathbb B = \mathbb D_N$. We shall
denote the mutual energy of interaction between these two subsets as $U(\mathbb
A, \mathbb B)$ for conciseness.  Furthermore, we define $m_\mathbb A$ to be the
sum of the elements of $\{m_i\}$ corresponding to the particles of $\mathbb A$.
Likewise, let $\vec v_\mathbb A$ denote the mass-weighted average velocity of
the elements of set $\mathbb A$. Our generalized definition of Hill's dimer to
the cluster is given below.
\begin{mydef}
\label{def:general}
The ordered set $\mathbb D_N$ of $N$ particles constitutes a physical cluster if
for all sets $\mathbb{A} \neq \emptyset$ and $\mathbb{B} = \mathbb D_N
\setminus \mathbb{A} \neq \emptyset$, we have that $\frac 1 2 \mu_\mathbb{AB}
(\vec v_\mathbb A - \vec v_\mathbb B)^2 \leq -U(\mathbb A, \mathbb B)$ where
$\mu_\mathbb{AB} = \frac {m_\mathbb A m_\mathbb B} {m_\mathbb A + m_\mathbb
B}$.
\end{mydef}

\section{Partition function of the Hill ensemble}
To treat the physical cluster in the theory of statistical mechanics, we need
to derive its partition function. For the sake of notational simplicity, we
shall concern ourselves only with the cluster at rest ($\vec v_\mathrm{cm} =
0$). Center-of-mass motion will be assumed separable from the internal degrees
of freedom. Hill \cite{hill55} gives the partition function for the dimer and
the explicit expression for the probability
\begin{equation}
p[-\beta U(\vec r_i, \vec r_j)] =  \mathrm{erf} \left (\sqrt{-\beta U(\vec r_i,
\vec r_j)} \right ) - \frac 2 {\sqrt{\pi}} e^{\beta U(\vec r_i, \vec r_j)}
\sqrt{-\beta U(\vec r_i, \vec r_j)}
\end{equation}
that two molecules a physical cluster at inverse temperature $\beta = 1 / kT$,
where $k$ is Boltzmann's constant and $T$ is the absolute temperature. Because
of the strong analogy between Definitions $\ref{def:dimer}$ and
$\ref{def:general}$, we are motivated to apply this probability function to
each term in Definition \ref{def:general}, but care must be exerted so as to
avoid double counting. We first consider the sets $\mathbb A_j$ that contain
the only element molecule $j = 1, 2, \ldots, N$. The probability that molecule
$j$ is part of the cluster is
\begin{equation}
p_j \equiv p[-\beta U(\mathbb A_j, \mathbb D_N \setminus \mathbb A_j)]
\end{equation}
Next we consider the sets $\mathbb A_{jk}$ that contain only the elements $j$
and $k$. The probability that the dimer $j,k$ is part of the cluster is
\begin{equation}
p_{jk} \equiv p[-\beta U(\mathbb A_{jk}, \mathbb D_N \setminus \mathbb A_{jk})]
\end{equation}
We continue to define probabilities like these, over greater and greater
subsets $\mathbb A_{jk\ldots}$ until we reach the subset that has $N / 2$
elements where we must stop to avoid double-counting.

The partition function for the cluster now follows upon substituting the
canonical probability function $e^{-\beta U(\{\vec r_i\})}$ by
$e^{-\beta U(\{\vec r_i\})} \prod_j p_j \prod_{j < k} p_{jk} \ldots$ in the
definition of the partition function. In the quantum case, we must deal with
the subtle issues raised by the Heisenberg uncertainty principle, but in the
classical case, this immediately gives
\begin{equation}
\label{eq:part}
Q_\mathrm{cm} = \frac {V} {N! \Lambda^{3 N}} \int \mathrm d \{\vec r_i\}
e^{-\beta U(\{ \vec r_i \})} \prod_j p_j \prod_{j<k} p_{jk} \prod_{j<k<l}
p_{jkl} \ldots
\end{equation}
This partition function in essence defines a new ensemble: the ensemble of the
physical cluster, or ``bound'' states. Its probability distribution is not
of Boltzmann form. We shall refer to this new ensemble as the ``Hill ensemble''
in his honor.

\section{Conclusion}
Definition \ref{def:general} is very general and encompasses both clusters of
molecules or atoms, or the molecules themselves. However, the physical cluster
is not defined for all potentials. For instance, the harmonic spring does not
satisfy the requirement that the potential energy vanish for large distances.
In physical reality, however, such an unbounded potential is never encountered,
and all molecules, and especially their clusters, are prone to disintegration
at high enough temperature. 

Because the products run over all dimers, trimers, $\ldots$, $N/2$-mers, the
numerical complexity in verifying whether $N$ molecules constitute a cluster
according to Definition \ref{def:general} is $(N/2)!$. Hence, numerical
experiments on the physical cluster are prohibitively expensive beyond the very
smallest ones.  However, generally it is clear that for $T \to 0$, we have
$p[-\beta U(\cdot)] \to 1$. Therefore, the Hill ensemble will reduce to the
canonical ensemble for low temperatures. This explains the observation by Lee,
Barker and Abraham \cite{lee73} that the thermodynamic properties of
Lennard-Jones clusters are largely independent of the choice of cluster
constraining radius at low temperature.

\bibliography{patron}
\end{document}